# Effect of $^{12}$C+$^{12}$C Reaction & Convective Mixing on the Progenitor Mass of ONe White Dwarfs

Ghina M. Halabi[a] & Mounib El Eid[a]

[a]American University of Beirut, Department of Physics, P.O. Box 11-0236, Riad El-Solh, Beirut, Lebanon, e-mail:gfm01@aub.edu.lb

**Abstract.** Stars in the mass range ~8 - 12 $M_\odot$ are the most numerous massive stars. This mass range is critical because it may lead to supernova (SN) explosion, so it is important for the production of heavy elements and the chemical evolution of the galaxy. We investigate the critical transition mass ($M_{up}$), which is the minimum initial stellar mass that attains the conditions for hydrostatic carbon burning. Stars of masses < $M_{up}$ evolve to the Asymptotic Giant Branch and then develop CO White Dwarfs, while stars of masses ≥ $M_{up}$ ignite carbon in a partially degenerate CO core and form electron degenerate ONe cores. These stars evolve to the Super AGB (SAGB) phase and either become progenitors of ONe White Dwarfs or eventually explode as electron-capture SN (EC-SN). We study the sensitivity of $M_{up}$ to the C-burning reaction rate and to the treatment of convective mixing. In particular, we show the effect of a recent determination of the $^{12}$C+$^{12}$C fusion rate, as well as the extension of the convective core during hydrogen and helium burning on $M_{up}$ in solar metallicity stars. We choose the 9 $M_\odot$ model to show the detailed characteristics of the evolution with the new C-burning rate.

## INTRODUCTION

According to the theory of stellar evolution, the majority of White Dwarfs (WDs) are post-AGB stars. The evolution to the AGB phase of stars less massive than a critical mass ($M_{up}$), leads to electron degenerate cores consisting of the ashes of He-burning, essentially carbon and oxygen, and develop CO WDs. Stars of M ≥ $M_{up}$ ignite carbon in a partially degenerate CO core and form electron degenerate ONe cores. After carbon burning, a star reaches the Super-AGB (SAGB) phase where hydrogen is re-ignited, recurrent thermal instabilities develop in the helium burning shell, and they become ONe WDs after losing their envelopes by stellar wind. SAGB stars may also explode as EC-SNe provided that their ONe cores have high enough central densities.

However, $M_{up}$ is not very well determined because of the uncertainties in the nuclear burning reaction rates, especially the $^{12}$C+ $^{12}$C fusion reaction rate, and the treatment of convective mixing. For stars of solar metallicity, $M_{up}$ ranges between ~6-9$M_\odot$ ([1], [2]). An up-to-date determination of $M_{up}$ and the abundance characteristic profiles of the WD is important for several reasons. It is important for the theory of novae outbursts in cataclysmic variables when these stars belong to close binary systems, as well as for SNIa explosions which require a CO WD growing to the Chandrasekhar mass, and thus are favored by a higher $M_{up}$.

$M_{up}$ depends directly on the masses of the He and CO cores which are highly dependent on the treatment of mixing for a given metallicity. Determining the extent of the convective region is still not settled. Classical models, i.e. those based on a bare Schwarzschild criterion are still used in many studies (e.g. [3]). Others assume that a diffusive overshooting due to the inertia of the convective elements takes place at the boundary of the convective region [4]. Including convective core overshooting during the H- and He-burning phases, is expected to lower the value of $M_{up}$. In the work by [5], the find that $M_{up}$ is $8.1 M_\odot$ for solar metallicity stars. However, they emphasize that this value is sensitive to the extension of the convective boundary of the core. In other works ([1], [6]), overshooting decreases $M_{up}$ from ~9 to $6 M_\odot$.

It is important to note that the value of this critical branching mass is also affected by other factors like rotation and binarity. Multi-dimensional simulations can possibly yield different values of $M_{up}$.

Given the current uncertainty in key reaction rates and convective mixing, it is important to investigate the effects of the $^{12}C+\,^{12}C$ fusion reaction rate on $M_{up}$ and on the composition profiles of the CO and ONe cores, as well as the effects of the extension of the convective core during hydrogen and helium burning.

## Status of the $^{12}C+\,^{12}C$ fusion reaction rate

The $^{12}C+\,^{12}C$ fusion reaction has two main channels:

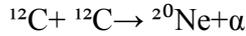
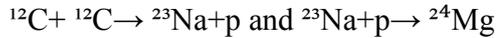

Thus, the most abundant species left over from core C-burning are $^{20}Ne$, $^{23}Na$, and $^{24}Mg$ in addition to $^{16}O$. This rate is very important not only for the C-burning phase but is also critical for SNIa, as well as for the mechanism powering superbursts in accreting neutron stars.

Despite considerable experimental efforts in the past decades, the total $^{12}C+\,^{12}C$ fusion reaction rate remains uncertain at stellar temperatures. Current data do not only show discrepancies in the derived astrophysical S-factor, but also in the possibility of having resonant structures, which makes the extrapolations to low energies extremely uncertain [7]. In our present models, the relevant temperature range is $T\approx 0.8\text{-}1.2 \times 10^9$ K which corresponds to $E_{cm} \approx 1.7\text{-}3.3$ MeV. In order to determine how the $^{12}C+^{12}C$ rate uncertainty affects $M_{up}$ and the composition profiles, we use 4 different $^{12}C+^{12}C$ rates: CF88 rate [8], the rate recommended by [7], as well as its upper and lower limits. We show in Fig.1 how these rates compare in the relevant temperature range.

The recommended rate (CR) is based on the classical extrapolation of the averaged S-factor data using a standard potential model plus the contribution of a single resonance observed at $E_{cm} = 2.13$ MeV [9], and is similar to the standard CF88 rate. The lower limit (CL) of the rate corresponds to the one suggested on the basis of incompressibility of nuclear matter in heavy ion fusion systematics by including a hindrance term for low-energy fusion processes [10]. The upper limit of the rate (CU) includes an additional term resulting from a possible strong $^{12}C+^{12}C$ cluster resonance at $E_{cm} = 1.5$ MeV.

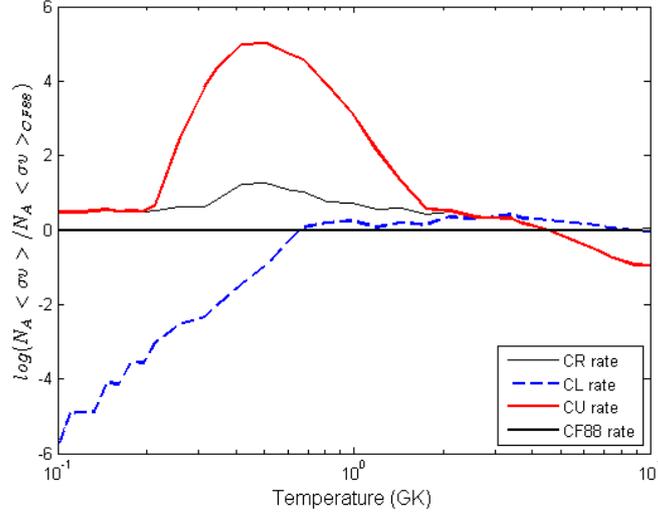

**FIGURE 1.** The $^{12}$C+ $^{12}$C reaction rates used in the present work. The rates are shown normalized to the CF88 rate. See text for details.

## Model Calculations

For solar metallicity stars, we find that with the CF88 rate, $M_{up}$= 8.5$M_\odot$. With the CR rate, the value is 8.3$M_\odot$. This is expected since CR is slightly more efficient than CF88 as Fig. 1 shows. With the CU rate, the value drops to 7$M_\odot$ because its efficiency is higher, while $M_{up}$ is 9$M_\odot$ with CL. Carbon ignites off-center with all the four rates used in the models of M= $M_{up}$.

We evolved the 9$M_\odot$ star through the carbon burning phase with the CR rate using the stellar evolution code described by ([11], [12]) and references therein. A major difference between carbon burning with the CF88 rate and the CR rate is that with the former, carbon ignites off-center, while with the CR rate, the ignition occurs centrally. This has important consequences on the shape of the abundance profiles at the end of carbon-burning. Fig. 2 shows the HR diagram and the evolution of the convective structure. The labels indicate relevant evolutionary phases in the two figures. We find that central carbon burning occurs in a convective core (see right panel of Fig. 2). After central carbon depletion, shell C-burning proceeds in zones surrounding the central core.

The abundance profiles at the end of carbon-burning are shown in Fig. 3. The figure shows a temperature inversion in the center due to neutrino energy losses which prevents carbon from getting completely depleted. This has significant consequences on novae eruptions.

Table 1 shows some characteristics of the He-exhausted (CO) core of the 9$M_\odot$ sequence at the end of carbon-burning, in the case of the CR rate. Note that $M_{CO}$ > 1$M_\odot$. The third most abundant nucleus in the ONe core is $^{23}$Na not $^{24}$Mg.

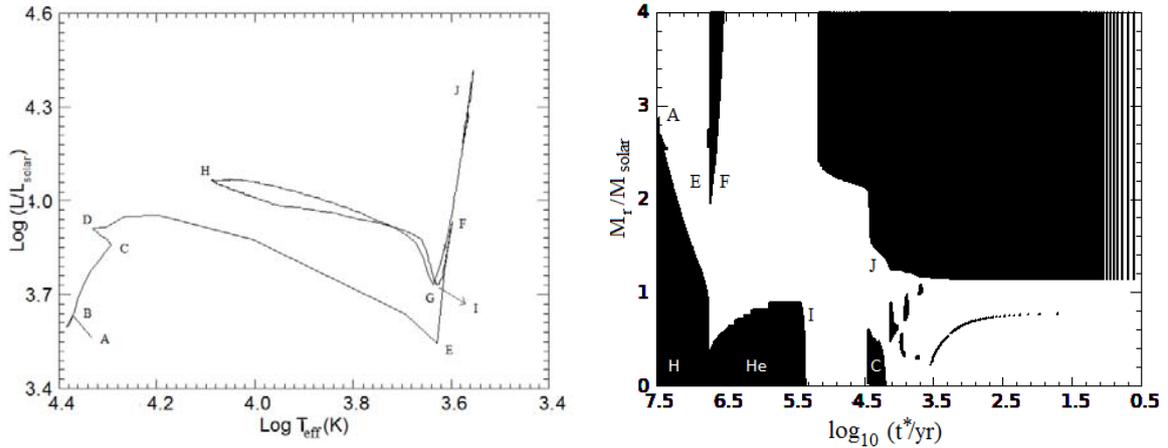

**FIGURE 2.** Left: Evolutionary track in the Hertzsprung-Russell diagram of a $9M_\odot$ sequence, of solar-like initial composition, from the ZAMS till the end of carbon burning. Different phases of evolution are labeled. Right: Evolution of the convective structure (black regions) for the H-, He- and C-burning phases of the $9M_\odot$ sequence. Labels are the same as those in the left panel. $t^*/yr$ is the time left until the end of calculation, i.e. $t^* = t_f - t$.

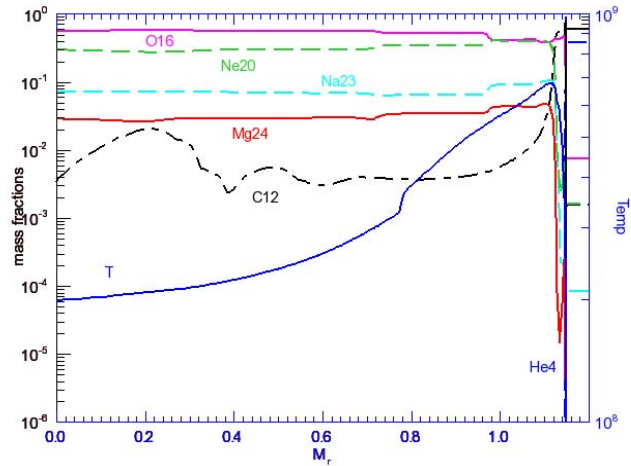

**FIGURE 3.** The abundance profiles of the important isotopes in the He-exhausted core of the $9M_\odot$ model at the end of core carbon burning with the CR rate.

**TABLE 1.** Core properties and central abundances of major isotopes at the end of carbon-burning with the CR rate for the $9M_\odot$ sequence. The quantities shown are the mass of the ONe core (the region in which the $^{12}C$ is almost completely depleted), the mass of the CO shell and the He-free core, central mass fractions, the maximum abundance by mass of the $^{12}C$ and its position and the abundance of $^1H$ and $^4He$ in the buffer zone on top of the H-exhausted core. Masses are in solar units.

| $M_{ONe}$ | $\Delta M_{CO}$ | $M_{CO}$ | $X_{12}^{max}$ | $M_r(X_{12}^{max})$ | $^{12}C_c$ |
|---|---|---|---|---|---|
| 1.122 | 0.025 | 1.147 | 0.021 | 0.215 | 0.0035 |
| $^{24}Mg$ | $^{16}O_c$ | $^{20}Ne_c$ | $^{23}Na_c$ | $^4He_{top}$ | $^1H_{top}$ |
| 0.029 | 0.57 | 0.299 | 0.0725 | 0.383 | 0.6 |

## Effect of core overshooting

In this section, we study the effect of core overshooting during H and He-burning phases on $M_{up}$. We use the mixing scheme described in [12]. The overshooting distance is expressed as $l=\gamma H_p$, where $H_p$ is the pressure scale height. In our recent attempt to constrain the extension of the convective overshoot during central H and He-burning, we find that the core overshooting distance should be confined to $0.1H_p$, as determined by our investigation on blue loops after the RGB phase [13]. Clearly, this overshoot leads to the growth of the convective core.

We find that with the CF88 rate, an extension of $0.1H_p$ lowers $M_{up}$ from $8.5M_\odot$ to $8M_\odot$, i.e. it shows a similar effect of a higher carbon fusion rate. In case of CR rate, $M_{up}$ is lowered from $8.5M_\odot$ to $7.5M_\odot$.

## CONCULSION

The new models obtained in this work show interesting carbon profiles in the ONe cores, with a carbon residual that may play a crucial role in novae outbursts, for example. This work is still in progress, aiming at studying the effect of the new abundance profiles on novae models (G. M. Halabi, J. Jose, M. El-Eid, in prepration). The new results are expected to improve SN models, given the uncertainties in their progenitor masses, as well as in their explosion mechanisms.

## ACKNOWLEDGMENTS

G.H. thanks the organisers of CSSP14 and the American University of Beirut (AUB) for support.